\documentclass[journal=nalefd,manuscript=letter,layout=traditional]{achemso}

\usepackage{graphicx}
\usepackage{dcolumn}
\usepackage{bm}
\usepackage{color}
\usepackage{xfrac}
\usepackage[normalem]{ulem}
\usepackage{cancel}
\graphicspath{ {./figures/} }
\usepackage{url}
\usepackage{mathtools}
\usepackage{hyperref}
\usepackage{caption}
\usepackage{physics}
\usepackage{xcolor}
\usepackage{booktabs}
\usepackage{multirow}
\usepackage{cuted} 

\newcommand{\cinam}{CNRS/Aix-Marseille Universit\'e, Centre Interdisciplinaire de Nanoscience de Marseille UMR 7325 Campus de Luminy, 13288 Marseille Cedex 9, France}

\newcommand{\UniSoochowPhys}{School of Physical Science and Technology, Soochow University, Suzhou 215006, People’s Republic of China}

\newcommand{\UniSoochowLab}{Jiangsu Key Laboratory of Frontier Material Physics and Devices, Soochow University, Suzhou 215006, People’s Republic of China}

\newcommand{\UniSuzhouOpt}{School of Optical and Electronic Information, Suzhou City University, Suzhou 215104, People’s Republic of China}

\newcommand{\UniSuzhouLab}{Jiangsu Key Laboratory and Suzhou Key Laboratory of Biophotonics, Suzhou City University, Suzhou 215104, People's Republic of China}

\newcommand{\ismmilano}{Istituto di Struttura della Materia-CNR (ISM-CNR) and ETSF, Piazza Leonardo da Vinci 32, 20133, Italy}

\newcommand{\ismmlib}{Istituto di Struttura della Materia-CNR (ISM-CNR), Montelibretti, Italy}

\title{Dark exciton signatures in the infrared transient absorption of MoS$_2$ monolayer}

\author{Tian-Xiang Qian}
\affiliation{\UniSoochowPhys}
\alsoaffiliation{\UniSoochowLab}

\author{Marco D'Alessandro}
\affiliation{\ismmlib}

\author{Claudio Attaccalite}
\affiliation{\cinam}

\author{Tian-Yi Cai}
\affiliation{\UniSoochowPhys}
\alsoaffiliation{\UniSoochowLab}
\email{caitianyi@suda.edu.cn}

\author{Sheng Ju}
\email{jusheng@szcu.edu.cn}
\affiliation{\UniSuzhouOpt}
\alsoaffiliation{\UniSuzhouLab}

\author{Davide Sangalli}
\email{davide.sangalli@cnr.it}
\affiliation{\ismmilano}

\begin{document}
	
	\begin{abstract}
	Dark excitons play a central role in the nonequilibrium dynamics of two-dimensional semiconductors, but remain difficult to characterize. Transient-absorption experiments, with probes tuned in exciton-exciton transitions energy range (exc-tr-abs), can detect excitations from any populated dark excitons, including symmetry-forbidden, spin-forbidden, and finite-momentum ones. In this work, we develop a $GW$+BSE scheme for computing exc-tr-abs spectra from arbitrary populated exciton distributions. These dark excitons are included equally by evaluating exciton-exciton dipoles in a locally smooth gauge, including intra- and inter-band contributions. For monolayer MoS$_2$, the exc-tr-abs signal arises from the $\Gamma$, $K$, $M$, and $Q$ valleys, differing substantially from the $\Gamma$-only interpretation. Exciton-exciton dipoles show similar intensities across these valleys, while their weights are dictated by initial excitonic populations. State- and spin-resolved analyses assign the peaks to $1s \rightarrow 2p$ and $1s \rightarrow 3p$ transitions from both spin-flip and spin-conserving A and B excitons across valleys and momenta.
	\end{abstract}

	
	Two-dimensional (2D) semi-conductors are promising candidates for novel opto-electronic devices, controlled by ultra short laser pulses and able to operate in the THz or even PHz regime.~\cite{Goel2021,Lemme2022} Among 2D materials, a special role is played by transition-metal dichalcogenides (TMDCs), with MoS$_2$ being the most studied case. TMDCs feature weak dielectric screening and a strong confinement effect that leads to unusually large exciton (electron–hole quasiparticle) binding energies, of the order of few hundred meV.~\cite{Wang2018} Excitons dominate the low-energy window (few eV) of their optical properties,~\cite{MolinaSanchez2013,Qiu2013a} and can be optically injected in TMDCs via laser pulses, by tuning the laser frequency below the electronic bandgap of the material.
	The excitons activated by the laser pulses are the bright ones, a very specific subset of the rich exciton-landscape available in 2D materials.~\cite{Wu2015,Qiu2015,Malic2018} Bright excitons have zero center of mass momentum ($\mathbf{q}=0$), same (opposite) spin between the valence-band (hole) and the conduction-band (electron),
	and finite transition dipole, $\bm{\mu}_{0\lambda}$, between the ground state and the excitonic state, enabling the coupling with light. In the Wannier picture, the envelope of the excitonic wave function can be described in terms of hydrogen-like states, while a more rigorous classification can be based on the material point group.~\cite{Bajaj2025,Nalabothula2026,Stohler2026} (see also S.I.6 of Supporting Information) Bright excitons belong to the ones with ``\textit{s}-like'' excitonic envelope. In the case of TMDCs, and in particular for MoS$_2$, the energy, lifetime and linewidth of bright excitons have been extensively characterized, both theoretically and experimentally, for free-standing samples.~\cite{Cadiz2017,Boule2020,Chan2023} Also, the role of strain, the impact of the dielectric environment, the role of encapsulation, and the effect of combining 2D layers in heterostructures and Moiré structures have been well characterized.~\cite{Shi2013,Thygesen2017,Mungu2017,Sevik2026,Reho2024,Chan2025}
	The dynamics of bright excitons, and
	their sensitivity to changes in the material properties, upon the action of an ultrashort laser pulse, has also been the focus of a very intense research activity.~\cite{DalConte2015,Pogna2016,Trovatello2020,Lloyd2021,Trovatello2022}
	However, right after the generation of bright excitons by means of a pulse (pump pulse), their non-equilibrium dynamics leads to the population of dark excitons.~\cite{Slobodeniuk2016,Selig2018,Christiansen2019,Wang2018,Merkl2019} Dark excitons are a much larger class, including symmetry-dark (envelope \textit{p}-like, \textit{d}-like, etc .. ), spin-forbidden (spin-flip excitons), and finite-$\mathbf{q}$ excitons.
	Even when the pump pulse is tuned above the electronic bandgap, the system, before relaxing back into the electronic ground state, will go towards metastable states where a thermal distribution of both bright and dark excitons is realized, with their population peaked at the energy minimum of the excitonic band structure.~\cite{Rustagi2018,Christiansen2019,Gosetti2025}
	
	Dark excitons are much less studied and characterized.
	The most direct way to measure dark excitons, is based on time-resolved angle-resolved photoemission spectroscopy (TR-ARPES). TR-ARPES can, in principle, provide a very comprehensive characterization of exciton dynamics.~\cite{Perfetto2016,Christiansen2019,Perfetto2020,Stefanucci2026,Wu2026} In practice, however, due to the complex apparatus needed, the very low signal to noise ratio in 2D materials, and intrinsic limits in the energy and momentum resolution of the experimental setup, characterization of dark excitons properties remains not easy, and studies are limited.~\cite{Madeo2020,Dong2021,Gosetti2025} On the other hand, optical measurements are much easier to perform with table-top setup; they usually provide better time and energy resolution, and could allow a complementary characterization of dark excitons. However, one needs to devise a way to detect dark excitons in optical experiments. 
	Lowest-energy dark excitons can be detected via photoluminescence or magneto-photoluminescence~\cite{Robert2020}, either because they are out-of-plane bright, or because they become visible despite very low oscillator intensity~\cite{Reho2024}.
	Recent experimental and theoretical developments exploited non-linear optical properties to study symmetry-forbidden, i.e. \textit{p}-like envelope, excitons~\cite{Ye2014,Wang2015,Panna2019,Qian2024,EsteveParedes2025,Montanaro2026}. In TMDCs, a detailed understanding of these experiments, requires a quantitative characterization of the energy difference between 1\textit{s}, 2\textit{s}, 3\textit{s}, etc... excitons, and the 2\textit{p}, 3\textit{p}, etc... counterparts. Already for the 1\textit{s} to 2\textit{p} energy difference, there are contrasting experimental results for different TMDCs~\cite{Poellmann2015,hill2015,Vaquero2020}. Most importantly, non-linear optics, just slightly broadens the range of excitons which can be directly explored, still remaining blind to many symmetry forbidden excitons, to spin forbidden (or spin-flip) excitons, and to finite-$\mathbf{q}$ excitons.
	
	An alternative approach, fully based on optical laser pulses, and able to detect transition involving virtually any dark excitonic state, is constituted by a setup where, after the initial excitation, either low energy photolumescence~\cite{Andrianov2019} or a second laser pulse (probe pulse) is used to probe transitions between excitonic states~\cite{Poellmann2015,Cha2016,Steinleitner2017,Steinleitner2018,Merkl2019,Merkl2020}. We refer to the latter as inter-excitons transient absorption, or exc-tr-abs in short. Exc-tr-abs in TMDCs requires probe pulses in the energy range of few tens of meV, which are becoming now accessible. Despite the absorption of the probe pulse is still limited by optical selection rules (symmetry constraints, no spin-flip, zero momentum transfer), dark excitons can be probed, because these constraints now apply to inter-exciton transition dipoles $\bm{\mu}_{\lambda\lambda'}$, between an initial and a final exciton state.~\cite{Sangalli2023} The initial excitonic state, which can be either bright or dark, has different properties compared to the ground state. Indeed a ``\textit{s}''-exciton can be excited into a ``\textit{p}'' exciton, a ``\textit{p}''-exciton into a ``\textit{d}''-exciton or back into an ``\textit{s}''-exciton, and so on. Also, a spin-flip exciton can be excited to another spin-flip exciton, and an exciton with finite-$\mathbf{q}$ can be excited to another exciton with the same $\mathbf{q}$.
	A very rich time-dependent signal can thus be measured experimentally, with all the information about the properties of bright and dark excitons encoded in it. Such richness naturally leads to a rather difficult interpretation of the experimental data, which have been so far interpreted in terms of 1\textit{s} to 2\textit{p} transitions at $\mathbf{q}=\Gamma$.~\cite{Poellmann2015,Cha2016,Steinleitner2018,Merkl2019,Merkl2020,Meineke2024} Accordingly, it calls for accurate theoretical and numerical simulations, able to analyze the experimental signal, to disentangle the contributions from different excitonic states, and eventually to guide future experiments.
	
	In the present work, building on top of previous developments,~\cite{Sangalli2023} we present a fully ab-initio scheme, based on state-of-the-art $GW$ plus Bethe-Salpeter Equation ($GW$+BSE), to model exc-tr-abs. In particular, we extend the numerical method in order to include finite-$\mathbf{q}$ excitons, and to capture intra-band physics in a gauge consistent way.
	Absorption spectra are defined by the dielectric tensor $\epsilon_{\alpha\beta}(\omega)$. 
	In the length gauge, it is written in terms of the dipole--dipole response function as
	${\epsilon_{\alpha\beta}(\omega) =
		\delta_{\alpha\beta} - 4\pi\chi_{\mu_\alpha\mu_\beta}(\omega).}$
	Here, $\alpha$ and $\beta$ denote Cartesian directions. Alternatively, the dielectric tensor can be formulated in the velocity gauge using the current--current response function. In this work, we employ the length-gauge formulation because the velocity gauge can suffer from numerical divergences when a finite basis set is used.\cite{Virk2007}
	At equilibrium, the absorption involves transitions between the ground state $|0\rangle$ and some zero momentum final excitonic state $|\lambda 0\rangle$. The dipoles entering the response function can be expressed as $\bm{\mu}_{0\lambda}$.
	In the non-equilibrium regime, an initial excitonic state $|\lambda_i\mathbf{q}\rangle$, with energy of $E_{\lambda_i}(\mathbf{q})=E_0+\omega_{\lambda_i}(\mathbf{q})$, can be populated. Here $E_0$ is the ground state energy. The subsequent probe pulse induces inter-exciton transitions from this initially populated state, and the corresponding dipole--dipole response can be written as~\cite{Sangalli2023}
	\begin{equation}
		\chi^{\lambda_i\mathbf{q}}_{\mu_\alpha\mu_\beta}(\omega)
		=
		\frac{2}{V} \sum_{\lambda}
		\frac{
			\mu^{\alpha}_{\lambda_i\lambda}(\mathbf{q})
			\mu^{\beta}_{\lambda\lambda_i}(\mathbf{q})
		}{
			\omega_{\lambda}(\mathbf{q})
			- \omega_{\lambda_i}(\mathbf{q}) - \omega - i\eta
		}
	\end{equation}
	$\bm{\mu}_{\lambda\lambda_i}(\mathbf{q})$ denotes the transition dipoles between the initial excitonic state and the final excitonic states $|\lambda\mathbf{q}\rangle$ reached by the probe pulse.~\footnote{The current–current response function can also be obtained in the same manner by replacing the dipole matrix elements $\mu^{\alpha}_{\lambda_i\lambda}(\mathbf{q})$ with the corresponding matrix elements $j^{\alpha}_{\lambda_i\lambda}(\mathbf{q})$.}
	For an optically injected exciton, one has to consider a zero momentum bright exciton state $|\lambda^{i},\mathbf{q}\rangle=|\lambda^{b},\bm{0}\rangle$ as the initial state.~\cite{Sangalli2023}
	On the other hand, if the system is probed with some delay with respect to the initial pump pulse, the initially excited population can relax and redistribute among different excitonic branches. In such case, the total dipole–dipole response is obtained by summing over all populated initial excitonic states,
	${
		\label{eq:population-averaged-chi}
		\chi_{\mu_\alpha\mu_\beta}(\omega)
		=
		\sum_{\lambda_i,\mathbf{q}}
		N_{\lambda_i}(\mathbf{q})
		\chi^{\lambda_i\mathbf{q}}_{\mu_\alpha\mu_\beta}(\omega).
	}$
	Here, the excitonic distribution $N_{\lambda_i}(\mathbf{q})$ weights the contributions from different initial excitonic states $|\lambda_i\mathbf{q}\rangle$. After relaxation, the population eventually leads to a thermal distribution of bright and dark excitons within the finite-momentum excitonic bands.
	This population is assumed to follow a Boltzmann distribution.~\cite{Katzer2023}
	Due to the summation 
	over the initial states,
	the exc-tr-abs signal is an average of signals including both dark and bright excitonic states at any finite-$\mathbf{q}$. In the calculation, all symmetry-equivalent $\mathbf{q}$-points in the Brillouin zone (BZ) must be explicitly included, since equivalent points can respond differently to the probe polarization.
	

	
	Although numerically stable, the length gauge involves a position operator that is ill-defined under periodic boundary conditions. Position dipoles are often reconstructed as $r_{nm\mathbf{k}}=v_{nm\mathbf{k}}/\Delta\epsilon_{nm\mathbf{k}}$, but this expression fails for intra-band or degenerate transitions. For exciton--exciton transitions, reconstructing $\mu_{\lambda\lambda_i}(\mathbf{q})$ from $v_{\lambda\lambda_i}(\mathbf{q})$ also requires an approximate velocity operator that neglects the electron--hole interaction kernel, introducing uncertainties in the dipole intensities (see S.I.1 and S.I.2). To avoid these limitations, we instead take advantage of a refined definition of the position dipoles, also including intra-band transitions~\cite{louielength}
	\begin{equation}
		\label{eq:length_dipole}
		\begin{aligned}
			\bm{\mu}_{\lambda\lambda_i}(\mathbf{q})
			=
			&\sum_{v,c\neq c',\mathbf{k}}
			A^{\lambda\mathbf{q}\,*}_{cv\mathbf{k}}
			A^{\lambda_i\mathbf{q}}_{c'v\mathbf{k}}
			\mathbf{r}_{cc'\mathbf{k}}
			-
			\sum_{c,v\neq v',\mathbf{k}}
			A^{\lambda\mathbf{q}\,*}_{cv\mathbf{k}}
			A^{\lambda_i\mathbf{q}}_{cv'\mathbf{k}}
			\mathbf{r}_{v'v\mathbf{k}-\mathbf{q}}
			+
			\sum_{cv,\mathbf{k}}
			A^{\lambda\mathbf{q}\,*}_{cv\mathbf{k}}
			\left( r_{cc\mathbf{k}}-r_{vv\mathbf{k}}  +i\partial_{\mathbf{k}} \right)
			A^{\lambda_i\mathbf{q}}_{cv\mathbf{k}}
		\end{aligned}
	\end{equation}
	$\mathbf{r}_{ll'\mathbf{k}}$ represents the Berry-connection matrix element associated with the Bloch states at $\mathbf{k}$. It includes both inter-band ($l\neq l'$) and intra-band ($l=l'$) contributions and is written as
	$
	\mathbf{r}_{ll'\mathbf{k}}=
	i\langle u_{l\mathbf{k}}|\partial_{\mathbf{k}}|u_{l'\mathbf{k}}\rangle,
	$
	where $|u_{l\mathbf{k}}\rangle$ denotes the cell-periodic part of the Bloch wave function.
	The terms 
	$
	A^{\lambda\mathbf{q}\,*}_{cv\mathbf{k}}
	\partial_{\mathbf{k}}
	A^{\lambda_i\mathbf{q}}_{cv\mathbf{k}}
	$
	account for another piece of intra-band physics, and constitute a generalization to the case $\lambda\neq\lambda'$ of the excitonic shift vector which has been recently studied.~\cite{Resta2024,Jiaming2026}
	In numerical \textit{ab initio} calculations, wave-function phases are arbitrary at each $\mathbf{k}$ point, and the derivatives $\partial_{\mathbf{k}}A^{\lambda_i\mathbf{q}}_{cv\mathbf{k}}$ and $\partial_{\mathbf{k}}|u_{l'\mathbf{k}}\rangle$ therefore require a consistent gauge. We make Equation~\eqref{eq:length_dipole} well-defined using a locally smooth gauge based on singular value decomposition (SVD), which ensures smooth excitonic amplitudes and cell-periodic Bloch functions between neighboring $\mathbf{k}$ points (see S.I.3 of Supporting Information).

	
	\begin{figure}[t]
		\includegraphics[width=0.45\textwidth]{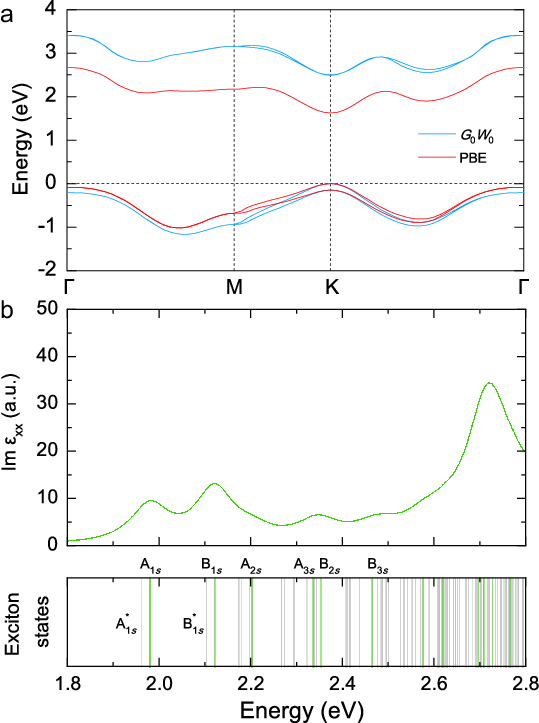}
		\caption{\label{fig:bands_and_abs} Equilibrium properties of monolayer MoS$_2$. (a) Electronic band structure at the $G_0W_0$ (blue solid curve) and PBE (red solid curve) approximations. (b) Optical absorption with electron-hole interaction considered. The excitonic states are shown in the bottom panel, including both bright (green lines) and dark (gray lines) excitons. The n\textit{s} states for A and B series are also labeled.}
	\end{figure}
	
	We compute exc-tr-abs in monolayer MoS$_2$, for which experimental data are available.~\cite{Cha2016} Its nonplanar three-atom unit cell has $D_{3h}$ symmetry without inversion.
	At Perdew-Burke-Ernzerhof (PBE) level, it is a direct bandgap semiconductor with a bandgap of 1.62 eV at $K$ point.
	The inclusion of spin-orbit coupling (SOC) strongly splits the valence band maximum (VBM) at $K$ point by 149 meV, whereas the conduction band minimum (CBM) exhibits a smaller splitting of 3 meV.
	Self-energy corrections at the $G_0W_0$ level significantly modify the band structure. 
	The bandgap remains direct with a value of 2.54 eV, where both the VBM and CBM unchanged, as shown in Figure~\ref{fig:bands_and_abs}a.
	Computational details are shown in S.I.4 of Supporting Information.
	
	To identify the bright exciton states relevant for the pump-probe response, let us quickly review the equilibrium absorption spectrum obtained from BSE, including both direct and exchange electron-hole interaction. As shown in Figure~\ref{fig:bands_and_abs}b, the absorption spectrum is dominated by excitonic peaks below bandgap.
	The two lowest energy peaks give rise to the so-called A (originating from the VBM) and B (originating from the VBM-1) excitons series.~\cite{MolinaSanchez2013,Qiu2013a} The lowest energy peak is due to two degenerate A$_{1s}$ bright excitons, due to transitions from VBM $\uparrow$ ($\downarrow$) to CBM+1 $\uparrow$ ($\downarrow$) at $K$ ($K$'). Similarly, two degenerate excitons, B$_{1s}$ are associated to the second peak, and they are due to transitions from VBM-1 $\downarrow$ ($\uparrow$) to CBM $\downarrow$ ($\uparrow$) at $K$ ($K$'). These bright excitonic transitions are spin-allowed, with the valence and conduction bands having the same spin. Moreover, two spin-forbidden excitons, denoted as A$^*_{1s}$ and B$^*_{1s}$, are located about 20 meV below the A$_{1s}$ and B$_{1s}$ bright excitons, respectively. For the other bright excitons, in order of increasing energy, we label them as A$_{2s}$, A$_{3s}$, B$_{2s}$, and B$_{3s}$.
	
	\begin{figure*}[!t]
		\includegraphics[width=0.8\textwidth]{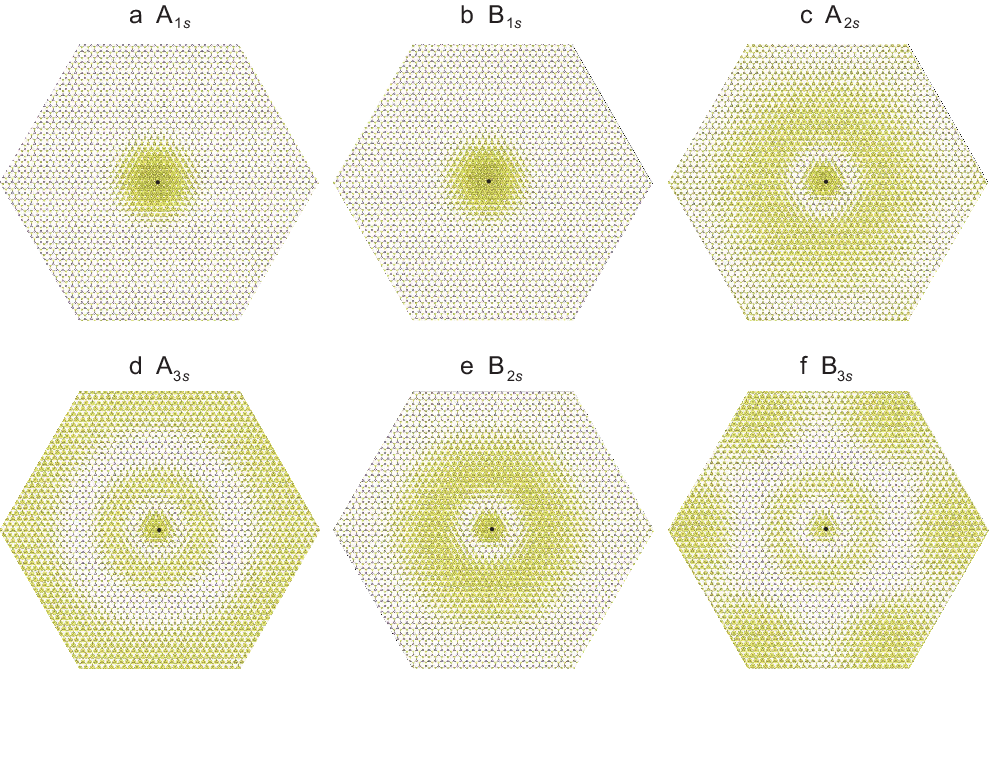}
		\caption{\label{fig:excitons-rspace}Real-space plots of the modulus squared of the exciton wave function for A and B n$s$ bright excitons. Here, the hole (black circle) is fixed at the Mo atom.}
	\end{figure*}
	
	The $\mathrm{n}s$ exciton series can be identified from the real-space exciton envelope functions shown in Figure~\ref{fig:excitons-rspace}, reflecting a hydrogen-like nodal structure. The $1s$ state is nodeless and strongly localized, the $2s$ state exhibits one radial node with a ring-shaped distribution, and the $3s$ state contains two radial nodes with more extended oscillatory behavior. 
	The figure shows that the labeling of excitons according to the (2D)-hydrogen model is well justified at $\mathbf{q}=\Gamma$. We adopt it here also for finite-$\mathbf{q}$ excitons, despite the picture is less clear in other valleys (see S.I.6 and S.I.7 of Supporting Information).
	
	\begin{table*}[b]
		\centering
		\caption{ Exciton energies in eV of A- and B-series n\textit{s} states in monolayer MoS$_2$.}
		\label{tab:s-excitons}
		\small
		\renewcommand{\arraystretch}{1.15}
		\setlength{\tabcolsep}{4pt}
		\begin{tabular}{lcccc}
			\toprule
			&
			Series &
			$\omega_{1s}$ [eV] &
			$\omega_{2s};\,(\omega_{2s}-\omega_{1s})$ [eV] &
			$\omega_{3s};\,(\omega_{3s}-\omega_{1s})$ [eV] \\
			\midrule
			
			\multirow{2}{*}{This work}
			& A & 1.98 & 2.20; (0.22) & 2.34; (0.36) \\
			& B & 2.12 & 2.35; (0.23) & 2.47; (0.35) \\
			\hline
			
			\multirow{2}{*}{Exp.\cite{hill2015}}
			& A & 1.88 & 2.05; (0.17) & 2.15; (0.27) \\
			& B & 2.03 & 2.24; (0.21) & 2.34; (0.31) \\
			\hline
			
			\multirow{2}{*}{Exp.\cite{Vaquero2020}}
			& A & 1.92 & 2.09; (0.17) & 2.13; (0.21) \\
			& B & 2.07 & 2.23; (0.16) & 2.30; (0.23) \\
			\hline
			
			\multirow{2}{*}{Exp.\cite{Xiaofeng2024}}
			& A & 1.82 & 1.98; (0.16) & 2.03; (0.21) \\
			& B & 1.92 & 2.13; (0.21) & 2.19; (0.27) \\
			
			\bottomrule
		\end{tabular}
	\end{table*}
	
	Because exciton energies and their splittings are sensitive to dielectric screening ~\cite{Ugeda2014,Raja2017,Wei2024}, we compare our free-standing MoS$_2$ results with equilibrium-absorption measurements on substrates before studying exc-tr-abs signal.~\cite{hill2015,Vaquero2020,Xiaofeng2024} Our A$_{1s}$ and B$_{1s}$ energies are approximately 0.1 eV higher than experimental values, whereas their separation of approximately 0.15 eV agrees with experiments and previous BSE calculations.~\cite{Qiu2013a} The calculated $2s$--$1s$ and $3s$--$1s$ splittings exceed experimental values \cite{hill2015,Vaquero2020} by a few tens of meV (Table~\ref{tab:s-excitons}).
	Likewise, $\omega_{\mathrm{A}_{2s}}-\omega_{\mathrm{A}_{1s}}$ of 0.28 eV \cite{Qiu2016} and 0.32 \cite{EsteveParedes2025} eV obtained in other theoretical calculations both exceed the experimentally measured 0.17 eV.
	We attribute these differences mostly to the dielectric screening from the substrate, which reduces the Coulomb interaction between electrons and holes, consequently leading to smaller energy separations between excitonic states with different principal quantum numbers.~\cite{Ugeda2014,Raja2017,Wei2024,Thygesen2017} The energy splitting $\omega_{\mathrm{A}_{2s}}-\omega_{\mathrm{A}_{1s}}$ is the most stable experimentally, with a shift of $\approx 50$ meV, compared to our results. We assume this as a reference value.
	
	
	\begin{figure*}[t]
		\includegraphics[width=0.8\textwidth]{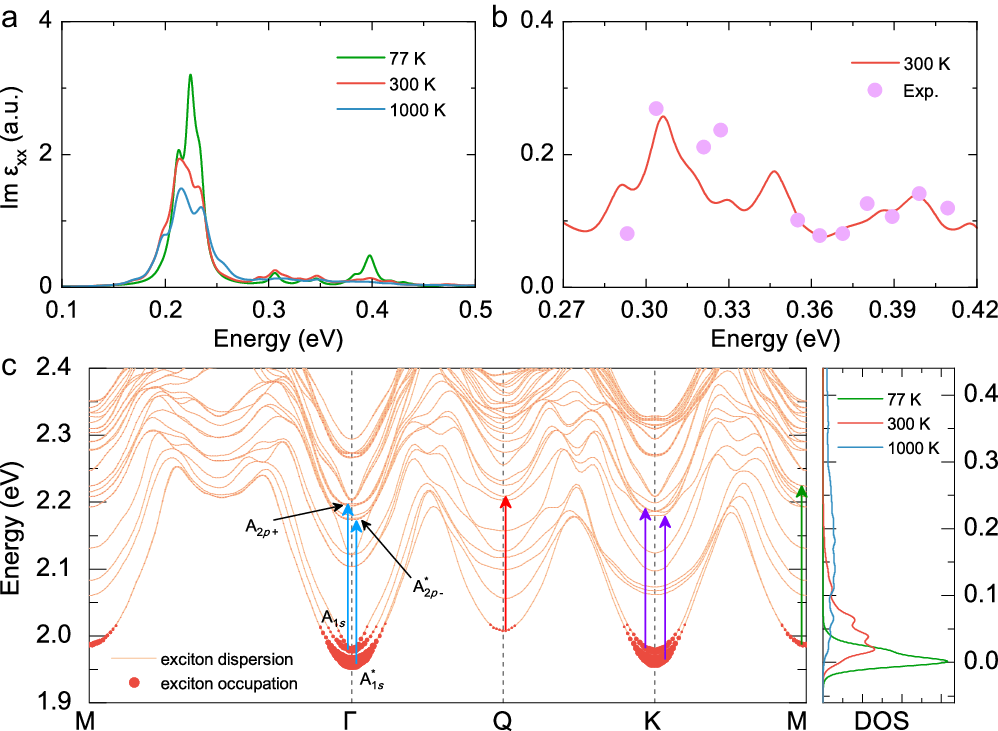}
		\caption{\label{fig:transient_abs_full}
			Temperature-dependent exc-tr-abs spectra of monolayer MoS$_2$.
			(a) Exc-tr-abs spectra in monolayer MoS$_2$ due to Boltzmann distributions of excitons at temperatures of $77$, $300$, and $1000$ K. (b) Zoom into the energy range 0.27-0.42 eV, where experimental pump--probe data are available (here  blueshifted by $46$ meV). (c) Exciton occupations (red dots) at 300 K represented on the exciton dispersion (left) and population weighted density of excitonic states (right) at $77$, $300$, and $1000$ K. 
			The exciton--exciton transitions contributing to the main peaks for every valley in Figure~\ref{fig:component}c are highlighted by arrows.
		}
	\end{figure*}
	
	After characterizing the equilibrium bright excitons through linear absorption, we next turn to the nonequilibrium exc-tr-abs spectra.
	In recent pump-probe experiments, a 3.1 eV pump pulse excites electron--hole pairs into high-energy continuum states.~\cite{Cha2016} These carriers relax through electron--phonon and electron--electron scattering, forming excitons across different branches and finite-$\mathbf{q}$ states. Before the probe arrives, the exciton population reaches quasi-equilibrium.
	The authors\cite{Cha2016} discussed the measured signal in terms of transitions between the bright A$_{1s}$ excitons and $p$ excitons at the  $\Gamma$ point.
	
	To check this description, here we compute exc-tr-abs by summing over a thermally populated excitonic distribution that includes contributions from both initial bright and dark excitons. 
	The exc-tr-abs spectrum in Figure~\ref{fig:transient_abs_full}a depends strongly on temperature: at $77$ K, a sharp peak around $0.22$ eV and a weaker shoulder near $0.24$ eV reflect strong exciton-exciton transitions from the lowest-energy states. At $300$ and $1000$ K, the main peaks broaden and weaken as finite-$\mathbf{q}$ populations increase. Weaker features around $0.30$--$0.40$ eV become more visible as thermally populated states from higher branches and momentum regions contribute to the exc-tr-abs response.
	Figure~\ref{fig:transient_abs_full}b compares the calculated exc-tr-abs spectrum at $300$ K with the experimental result at a delay of $0.9$ ps \cite{Cha2016}. The experimental data are blue-shifted by $\approx 50$ meV, corresponding to the $2s$--$1s$ energy separation discussed above. Specifically we refine the value to $46$ meV by aligning the energy position of the most intense peak. 
    Overall, the agreement in the main spectral features indicates that the thermally populated finite-$\mathbf{q}$ excitonic states provide a reasonable description of the observed exc-tr-abs response.
	
	To further clarify the origin of the exc-tr-abs features, we disentangle the contributions from different initial excitonic states. In Figure~\ref{fig:transient_abs_full}c, the exciton dispersion is plotted together with the Boltzmann occupation at an effective exciton temperature of $300$ K. The occupied excitonic states are mainly distributed around the $\Gamma$ and $K$ valleys, with additional contributions from the $M$ valley and a small occupation around the $Q$ valley.
	Accordingly, the exc-tr-abs signal is divided into four components, corresponding to excitonic states around the $\Gamma$, $K$, $M$, and $Q$ valleys.  For each valley, 18 or 24 full-BZ points generated by symmetry from the three or four most populated irreducible points are considered.
	The density of thermally populated excitonic states (DOS), shown in right panel of Figure~\ref{fig:transient_abs_full}c, evolves from a narrow peak near $1.96$ eV at $77$ K, dominated by the $\mathrm{A}_{1s}$ and $\mathrm{A}_{1s}^{*}$ states around $\Gamma$ and $K$, to a broader $1.95$--$2.05$ eV distribution at $300$ K as the $M$ and $Q$ valleys become populated, and becomes nearly flat at $1000$ K.
	

	\begin{figure*}[t]
		\includegraphics[width=1\textwidth]{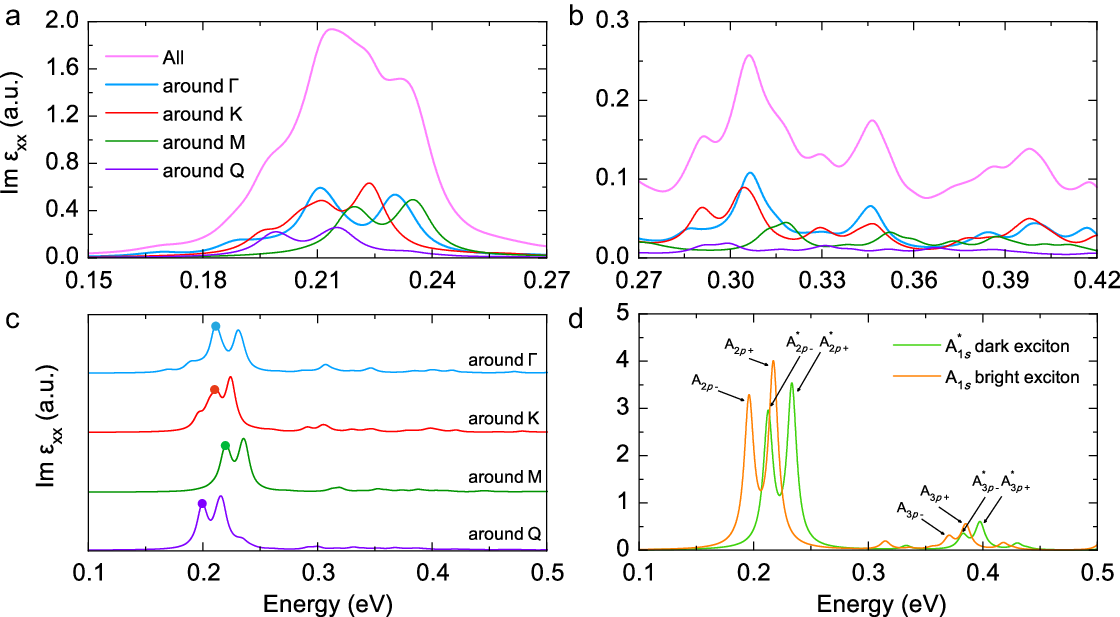}
		\caption{\label{fig:component} 
			Analysis of the exc-tr-abs spectrum in monolayer MoS$_2$ at an effective exciton temperature of $300$ K.
			(a,b) Population-weighted, momentum-resolved decomposition, in two different energy ranges, from the $\Gamma$ (blue), $K$ (red), $M$ (green), and $Q$ (purple) valleys. The total response is also represented (magenta).
			(c) Population independent valley contributions. The main peaks, marked by dots, correspond to the arrows in Figure~\ref{fig:transient_abs_full}.
			(d) Single state contribution, obtained by selectively populating the $\mathrm{A}^{*}_{1s}$ dark exciton, and the $\mathrm{A}_{1s}$ bright exciton at the $\Gamma$ point. The $p$ states are labeled for the main peaks.
		}
	\end{figure*}
	Figures~\ref{fig:component}a and \ref{fig:component}b show the resulting momentum-resolved decomposition of the exc-tr-abs spectrum. The $0.20$--$0.24$ eV main structure combines contributions from different valleys: $\Gamma$ dominates near $0.21$ eV, $K$ and $M$ at $0.22$--$0.24$ eV, and $Q$ contributes weak peaks near $0.19$ and $0.21$ eV. The weaker, more structured higher-energy features arise primarily from $\Gamma$ and $K$, with weak $M$ and negligible $Q$ contributions.
	Without population weighting, the four valleys show comparable intensities (Figure~\ref{fig:component}c), confirming that the stronger $\Gamma$ and $K$ contributions in Figure~\ref{fig:transient_abs_full}a arise from their larger populations.
	
	To resolve individual transitions, we examine contributions from selected excitonic states at fixed $\mathbf{q}$. At $\Gamma$, the dark $\mathrm{A}^*_{1s}$ and bright $\mathrm{A}_{1s}$ spectra have similar intensities and shapes, differing mainly by an energy shift (Figure~\ref{fig:component}d).
	%
	The main peaks originate from a pair of transitions, $\mathrm{A}_{1s}\rightarrow \mathrm{A}_{2p+}$ and $\mathrm{A}_{1s}\rightarrow \mathrm{A}_{2p-}$ in the bright channel, and two corresponding $\mathrm{A}^{*}_{1s}\rightarrow \mathrm{A}^{*}_{2p-}$ and $\mathrm{A}^{*}_{1s}\rightarrow \mathrm{A}^{*}_{2p+}$ transitions in the spin-flip channel. The pair is due to the splitting of the $p_\pm=p_x\pm p_y$ states.~\cite{Srivastava2015,Zhou2015,Yong2019} 
	The $\mathrm{A}_{1s}$ and $\mathrm{A}^{*}_{1s}$ initial states are split by $19$ meV, consistent with the experimental value of $14$ meV.~\cite{Robert2020} In contrast, the corresponding $\mathrm{A}_{2p-}$ and $\mathrm{A}^{*}_{2p-}$ final states are separated by only $\approx 2$ meV. The difference between the initial- and final-state splittings therefore results in an energy shift of $\approx 17$ meV between the two pairs of exc-tr-abs peaks. Moreover the $\mathrm{A}_{2p-}$ and $\mathrm{A}_{2p+}$ states are also split by $\approx 20$ meV, also in good agreement with recent SHG measurements~\cite{Qian2024}, where only the spin conserving doublet is detected. As a consequence the two pairs of peaks merge, leading to a three-peak structure at 300 K which is visible in Figure~\ref{fig:component}c for the $\Gamma$ valley. A similar structure is present in the $K$ valley, where however the peaks are mostly determined by the $\mathbf{q}$ close to $K$, rather than directly transitions at $\mathbf{q}=K$ (see S.I.5 of Supporting Information), likely due to the exciton dispersion shape of the $\mathrm{A}_{2p-}$ state, which flattens nearby $K$ (see Figure~\ref{fig:transient_abs_full}c). For the $Q$ and $M$ valleys instead the lowest energy state is not split into a same--spin and a spin--flip state, 
	and a two-peak structure remains. In Figure~\ref{fig:transient_abs_full}c, the main peak from the lowest energy state is marked by a dot, and the corresponding transition in Figure~\ref{fig:component}c is represented by arrows.
	(In S.I.5 of Supporting Information we further discuss how the peak energy positions result from an average of transitions from different $\mathbf{q}$-points within the $\Gamma$ valley.)

	In the high-energy region (Figure~\ref{fig:transient_abs_full}b), the authors of Ref. ~\cite{Cha2016}, by fitting the experimental data, identify three main spectral features, which they attribute to transitions between excitons with $\mathbf{q}=\Gamma$, as follows: $\approx0.32$ eV ($\mathrm{A}_{1s} \rightarrow \mathrm{A}_{3p}$), $\approx0.36$ eV ($\mathrm{B}_{1s} \rightarrow \mathrm{B}_{3p}$), and $\approx0.41$ eV ($\mathrm{A}_{1s} \rightarrow \mathrm{B}_{2p}$).~\footnote{We report here the values already including the shift of 46 meV.} 
	Instead, our calculations at an effective excitonic temperature of 300 K show that the signal is equally due to transitions from thermally populated excitons with $\mathbf{q}$ around the $\Gamma$ and $K$ valleys, while the $M$-valley contribution remains relatively weak. Remarkably, we also obtain 3 main features at $\approx0.31$ eV, $\approx0.35$ eV, and $\approx0.40$ eV, in excellent agreement with the experimental fit.
	However, in our calculations, all three features originate predominantly from $\mathrm{A}^{*}_{1s}\rightarrow \mathrm{A}^{*}_{3p}$ inter-exciton transitions at different $\mathbf{q}$-points nearby the $\Gamma$ and $K$ valleys.
	Specifically, the first two features at approximately $0.31$ and $0.35$ eV are dominated by $\mathrm{A}^{*}_{1s}\rightarrow \mathrm{A}^{*}_{3p}$ transition for $\mathbf{q}$ slightly away from $\Gamma$ and slightly away from $K$ (see Section S.I.5 of Supporting Information). The third feature at approximately $0.40$ eV instead mainly originates from $\mathrm{A}^{*}_{1s}\rightarrow \mathrm{A}^{*}_{3p}$ transition at $\mathbf{q}=\Gamma$ and $\mathbf{q}=K$.
	In addition, the transitions $\mathrm{B}^*_{1s} \rightarrow \mathrm{B}^*_{np\pm}$ and $\mathrm{B}_{1s} \rightarrow \mathrm{B}_{np\pm}$, with $n=2,3$, also exist in our simulations, but they give negligible signal due to the much lower population of the $\mathrm{B}$ excitons.
	Overall, most peaks on the exc-tr-abs spectra can be identified as belonging to the $\mathrm{X}_{1s}\rightarrow \mathrm{X}_{np\pm}$ and $\mathrm{X}^*_{1s}\rightarrow \mathrm{X}^*_{np\pm}$ series, with $\mathrm{X}=\mathrm{A,B}$ and $n=2,3$, and with contributions from different $\mathbf{q}$-points. Accordingly, the fine structure of the $\mathrm{A}_{1s}\rightarrow \mathrm{A}_{2p}$ peak at $\approx 0.2$ eV discussed above, with a double energy splitting, is also replicated at higher energy.
	Indeed, the exc-tr-abs spectra obtained by selectively populating
	the $\mathrm{A}^{*}_{1s}$ and $\mathrm{B}^{*}_{1s}$ excitons are
	nearly identical, as are those obtained from the $\mathrm{A}_{1s}$ and
	$\mathrm{B}_{1s}$ excitons (see Section S.I.5 of Supporting Information, Figure S4). No discernible signatures of transitions
	between the A and B series are observed. This also indicates that the
	transitions mainly involve changes in the conduction-band state of
	the electron, while the hole, which determines the A-B splitting retains its original valence-band character.
	

	
	In conclusion we have shown a fully \emph{ab initio} scheme, based on $GW$+BSE calculations, to model exc-tr-abs.
	The method includes finite-$\mathbf{q}$ excitons, and evaluates length--gauge exciton--exciton transition dipoles, including intra-band contributions in a locally smooth gauge.
	The overall exc-tr-abs signal is due to contributions from exciton-exciton transitions in different valleys in the exciton dispersion.
	%
	Applied to monolayer MoS$_2$, it captures the main relative features observed experimentally. 
	Our approach makes it possible to disentangle the contributions from the four valleys hosting local minima, namely $\Gamma$, $K$, $M$, and $Q$, as well as contributions from different transitions within each valley.
	Normalized valley- and spin-resolved spectra indicate that their relative contributions are largely controlled by the thermal populations, while transition dipole matrix elements are similar across different valleys. The main peak in the exc-tr-abs spectra emerges from a superposition of two pairs of peaks due to 
	A$^{*}_{1s} \rightarrow$ A$^{*}_{2p\pm}$ and A$_{1s} \rightarrow$ A$_{2p\pm}$ transitions at the $\Gamma$ and $K$ valleys. These are however superimposed to A$_{1s} \rightarrow$ A$_{2p\pm}$ at the $Q$ and $M$ valleys.
	At higher energies, where experimental data are available, the signal is instead dominated by A$^{*}_{1s} \rightarrow$ A$^{*}_{3p\pm}$ transitions at the $\Gamma$ and $K$ valleys, with an interpretation of the final signal substantially different from the one provided in the past, which was based on transitions at $\mathbf{q}=\Gamma$ alone.
	%
	%
	%
	These results show that indeed exc-tr-abs spectroscopy can be used as a sensitive probe for both dark and bright excitons dynamics, but also that the interpretation of the rich emerging signal needs to be complemented by accurate modelling.
	

	\section{Associated content}
	\textbf{Supporting Information}\\
	The Supporting Information is available free of charge.\\
	S.I.1. Exc-tr-abs spectra from velocity-derived and direct position dipoles, \\
	S.I.2. Sum--over--states vs time--dependent Berry phase at $\mathbf{q}=\Gamma$ in hBN monolayer, \\
	S.I.3. Locally smooth gauge through single value decomposition, \\
	S.I.4. Computational details, \\
	S.I.5. Analysis of the exc-tr-abs features near the $\Gamma$ and $K$ valleys, \\
	S.I.6. Exact symmetry analysis, \\
	S.I.7. Real-space plots of excitons with $\mathbf{q}=K$.

	\begin{acknowledgement}
		D.S. and C.A. acknowledge support from the European Union’s Horizon Europe research and innovation programme under the Marie Sklodowska-Curie grant agreement 101118915 (project TIMES). D.S. acknowledges funding from the PRIN project "Exploring extreme ultraviolet excitons with attosecond time resolution" (EXATTO), Grant No. 2022PX279E from MIUR (Italy), and MaX ``MAterials design at the eXascale” project, co-funded by the European High Performance Computing Joint Undertaking (JU) and participating countries
		(Grant Agreement No. 101093374).
		S.J. acknowledges funding from National Natural Science Foundation of China under Grant No. 52572321, Suzhou Basic Research Project (SJC2023003),
		and Wenzheng Scholarship of Suzhou City University.
	\end{acknowledgement}
	
	\bibliography{biblio}

\end{document}